# Colorectal cancer survival prediction using deep distribution based multiple-instance learning

Xingyu Li[1], Jitendra Jonnagaddala[2*], Min Cen[1], Hong Zhang[1*], Xu Steven Xu[3]*

[1] Department of Statistics and Finance, School of Management, University of Science and Technology of China, Anhui, P.R.China

[2] School of Population Health, UNSW Sydney, Kensington, Australia.

[3] Clinical Pharmacology and Quantitative Science, Genmab Inc., Princeton, New Jersey, USA.

* Corresponding author

**Abstract:**

Most deep-learning algorithms that use Hematoxylin and eosin stained whole slide images (WSIs) to predict cancer survival incorporate image patches either with the highest scores or combination of both highest and lowest scores. In this study, we hypothesize that incorporating wholistic patch information can predict the CRC cancer survival more accurately. As such, we developed a distribution based multiple-instance survival learning algorithm (DeepDisMISL) to validate this hypothesis on two large international CRC WSIs datasets called MCO CRC and TCGA COAD-READ. Our results suggest that combining patches that are scored based on the percentile distributions together with the patches that are scored as highest and lowest drastically improves the performance of CRC survival prediction. Including multiple neighborhood instances around each selected distribution location (e.g., percentiles) could further improve the prediction. DeepDisMISL demonstrated superior predictive ability compared to other recently published, state-of-the-art algorithms. Furthermore, DeepDisMISL is interpretable and can assist clinicians in understanding the relationship between cancer morphological phenotypes and patient's cancer survival risk. DeepDisMISL algorithm is available at https://github.com/1996lixingyu1996/DeepDisMISL

# 1 Introduction

Traditional risk stratification in cancer patients is usually based on cancer staging, grading, and molecular and clinical characteristics[1-3]. Tumor grade based on cellular appearance has been shown to be a measure of prognosis and an indicator of how quickly a tumor is likely to grow and spread[4]. However, high inter-observer variability has been a major challenge [5]. With the advances in the computer vision, deep learning algorithms have been successfully applied on Hematoxylin and Eosin (H&E) stained whole slide images (WSIs) to predict patient outcomes such as, overall survival, progression free survival, time to metastasis, or tumor recurrence. Additionally, WSIs were also used to stratify patients according to their survival risk [6-8].

WSIs were usually split into small image patches to train neural networks to predict an outcome of interest. Due to the large number of image patches for each WSI that can vary from hundreds to thousands, and unavailability of patch-level labels, it is not feasible to directly build a prediction model at patch level. Multiple-instance learning (MIL), a form of weakly supervised learning with only coarse-grained labels, can be used to predict outcomes at patient level [9]. For MIL, identification of WSI patches with the most predictive ability is important. Evidence suggests that patches with highest scores often carry the most predictive information for prediction. Recently, Courtiol et al. proposed MesoNet for survival data [8] which demonstrated that using both the lowest scored patches and the highest scoring patches can more accurately predict risk and survival in malignant mesothelioma patients.

In this study, we hypothesize that incorporating wholistic patch information can predict the CRC cancer survival more accurately. Therefore, we developed a distribution based multiple-instance survival learning algorithm (DeepDisMISL) to validate this hypothesis. DeepDisMISL is also highly interpretable and can assist clinicians in identification of morphological phenotypes. Using a 5-fold cross validation on a large Australian colorectal cancer dataset – MCO CRC [10, 11], and external validation on the TCGA COAD-READ dataset [12], we demonstrate that, by gradually incorporating patches scored and stratified into different percentiles (e.g., $1^{st}$, $5^{th}$, $25^{th}$, $75^{th}$, $95^{th}$, and $99^{th}$ percentiles) to the patches with highest and lowest score, the predictive performance of CRC survival prediction can be improved drastically. In addition, we systematically compared DeepDisMISL with six different baseline algorithms. The proposed DeepDisMISL demonstrates superior and robust predictive ability compared to all 6 baseline algorithms, including the recently published state-of-the-art algorithms like, MesoNet and DeepAttnMISL [6].

## 2. Methods

### 2.1 Datasets

The analysis included two large-scale datasets. WSIs (40x) were collected from both MCO CRC and TCGA CRC studies. The MCO CRC dataset was made available through the SREDH Consortium (www.sredhconsortium.org). The MCO CRC dataset consisted of patients who underwent curative resection for colorectal cancer between 1994 to 2010 in New South Wales, Australia[10, 11]. The Cancer Genome Atlas (TCGA) public dataset included the TCGA-COAD and TCGA-READ datasets. The MCO CRC dataset

(v15Jan2021) was used to train the deep learning model while the TCGA COAD and READ dataset obtained in July 2020 was used for external validation. For patients with more than one WSI available, we randomly selected one slide for each patient. Images with annotation marks or blur were excluded. After exclusion, 1184 WSIs from 1184 patients MCO patients were included in the analysis, whereas 529 WSIs from 529 patients were available from the TCGA database.

## 2.2 Preprocessing of WSIs

WSIs were first preprocessed to exclude the background area of each image where no tissue was present. OTSU algorithm was implemented to classify the image into two classes: the foreground (containing the matter) and background [13]. After removing the backgrounds, WSIs were cropped to non-overlapping, fixed-size tiles of images (224 * 224 pixels, 0.5mpp), which were then color normalized with Macenko's method. Depending on the size of the image, the number of tiles varied from a few hundred to 50,000 with a median of 12,047.

### 2.3 Feature extraction

We extracted features using a fine-tuned Xception model [14]. This neural network allowed us to obtain 256 relevant features from each tile. For each WSI (patient), a feature matrix with a dimension of n (number of tiles) x 256 (features) was obtained. We randomly selected 12,000 tiles from the feature matrix of each patient. If the tile number per slide was smaller than 12,000, we would resample the feature matrix up to the size of 12,000.

### 2.4 DeepDisMISL

Figure 1 shows the proposed DeepDisMISL algorithm. We incorporated MIL before predicting on the unlabeled tiles. WSIs were considered the bags of MIL (each WSI can be considered as a bag, and patches of the WSI can be seen as instances), for which supervision was provided only for the bag at individual patient level (i.e., survival time **t** and survival status **δ** for a patient), and the individual label of the instances contained in the bags (i.e., tiles) were not available. Two convolution one-dimensional (1-D) layers with ReLU activation were used to aggregate all the local features of the same tile into a global feature and devised a score (the element in the vector of 12000 * 1, each element can be seen as a score, Figure 1 and Table 2) for each tile. The scores at the selected percentiles (instances) of the global feature vector were the input features for the next prediction layer, which consisted of a two-layer multi-layer perceptron (MLP) classifier with two fully connected layers (128 and 64 neurons) and a ReLU activation for predicting survival risk with a cox loss function to deal with the censored survival data. The log likelihood function of the cox loss function used is equation 1. Where $O_i$ is the risk score, $\delta_i$ is the censoring variable ($\delta = 0$, death was not been observed). The neural network minimized the negative log partial likelihood function and produced maximum partial likelihood estimates.

$$L(O_i) = \sum_i \delta_i(-O_i - \log \sum_{j:t_j \geq t_i} \exp(O_j))$$

Equation 1: Log likelihood function of the cox loss function

We started with top/bottom instances (i.e., minimum, and maximum patch scores, or 0 and 100[th] percentile), and incrementally added additional instances at different percentiles to the top/bottom instances. In total, 7 different scenarios were investigated, i.e.,

(Scenario 1) [0, 100%],

(Scenario 2) [0, 0.1%, 99.9%, 100%],

(Scenario 3) [0, 0.1%, 1%, 99%, 99.9%, 100%],

(Scenario 4) [0, 0.1%, 1%, 5%, 95%, 99%, 99.9%, 100%],

(Scenario 5) [0, 0.1%, 1%, 5%, 10%, 90%, 95%, 99%, 99.9%, 100%],

(Scenario 6) [0, 0.1%, 1%, 5%, 10%, 25%, 75%, 90%, 95%, 99%, 99.9%, 100%], and

(Scenario 7) [0, 0.1%, 1%, 5%, 10%, 25%, 50%, 75%, 90%, 95%, 99%, 99.9%, 100%].

Zhu et al. showed that increasing number of top instances could improve the model performance. We also evaluated whether the number of neighborhood instances around each percentile has impact on the model predictive ability. In this experiment, we took the model with the most complete distribution (Scenario #7) and explored different number of neighborhood instances at each percentile (i.e., 1, 3, 5, and 7).

**2.5 Baseline algorithms**

We compared our algorithm with following different baseline algorithms (Figure 1): MesoNet [19], Meanpooling and Maxpooling (Top 1 and Top 10 instance), MeanFeaturePool and DeepAttnMISL [6]. Similar to DeepDisMISL, MesoNet used two 1-D convolution layers to aggregate all local feature descriptors of a tile into a global feature vector. Then, the largest and smallest 10 scores from the global feature vector were selected as the features for the classifier in the prediction layer. Similarly, Meanpooling and Maxpooling (Top 1 instance) used the mean score and max score, respectively, from the global feature vector generated by 1-D convolution layers as the features for the classifier in the prediction layer. Maxpooling (Top 10 instance) used the largest 10 scores from the global feature vector as the features for the classifier in the prediction layer. For Meanpool (Lasso cox), the features values for all tiles (12, 000) from a feature matrix (12000 x 256) for each WSI was averaged to obtain an average feature vector (1 x 256) for each WSI or patient. Lasso cox implemented in R package glmnet [15] was used to fit the survival data. The C-index values from a 5-fold cross validation were obtained from [13]. Both DeepAttnMISL [13] and our proposed DeepDisMISL used MCO CRC dataset for cross validation, and the C-index values were compared directly. We further compared the performance of DeepDisMISL with the other baseline algorithms by examination of their ability for risk stratification. For each model, the median risk score in the training set was calculated and then applied as a threshold to stratify each patient into high-risk or low-risk group.

## 2.6 Annotation of WSI patches

To interpret the model, we annotated the patches near the different percentiles of the patch scores. Kather et al. developed a deep-learning classifier to classify CRC image tiles into eight tissue types: adipose tissue (ADI), background (BACK), debris (DEB), lymphocytes (LYM), mucus (MUC), smooth muscle (MUS), normal colon mucosa (NORM), cancer-associated stroma (STR), and colorectal adenocarcinoma epithelium (TUM) [16]. We used pathologist annotated NCT-CRC-HE-100K and CRC-VAL-HE-7K image sets provided by Kather et al. to train and validate, respectively, a similar tissue-type classifier. We downloaded the Xception model from Keras and fine-tuned the model using the NCT-CRC-HE-100K set to develop the tissue-type classifier[14]. The overall accuracy of the tissue-type classification model was 99% based on training dataset, NCT-CRC-HE-100K and 94.4% based on the validation image set, CRC-VAL-HE-7K (Supplementary Figure 1). The tissue type of each image tile from the TCGA dataset was predicted using the fine-tuned Xception-based tissue-type classifier.

## 2.7 Evaluation

All models were trained using 5-fold cross validation on the MCO CRC dataset. In each fold, 80% of the data were used for model training and 20% of the data were used for model validation. For training, we used Adam optimization with grid search strategy. The training process monitored the loss on the MCO validation dataset and was designed to stop if the loss increased goes increased much. We evaluated the model performances with concordance index (C-index) in the survival prediction task. The TCGA data served as the independent, external validation dataset.

## 3 Results

### 3.1 Evaluation of probability distribution-based patch selection

Figure 2a shows that, based on the 5-fold cross validation using the MCO CRC dataset, there is an obvious trend where the more percentiles, the higher the C-index. With the top/bottom instances (See Figure 1 c), the average C-index was 0.611 (range: 0.58 – 0.630). Adding 2 more percentiles at 0.1% and 99.9% (Scenario #2) improved the average C-index to 0.62 (0.59 – 0.638). As expected, Scenario #7 with the most complete distribution information (0, 0.1%, 1%, 5%, 10%, 90%, 25%, 50%, 75%, 95%, 99%, 99.9%, 100%) produced the best predictive performance with an average C-index of 0.638 (0.626 – 0.66). The increasing trend in C-index suggests that the complete distribution of the patch scores carries richer information of the WSI than the only top and bottom instances (Figure 3).

Similarly, the external validation using an independent TCGA COAD-READ dataset demonstrated a similar trend (Figure 2b). In general, the models with more complete distribution (i.e., more middle scoring patches at different distribution percentiles in addition to the highest/lowest scoring patches) provides higher C-index compared to those with less complete distribution (e.g., only the highest and lowest scoring patches). As expected, the external validation with independent dataset produced more heterogeneous results and lower C-index compared to the cross validation since the TCGA dataset does not completely resemble the MCO CRC dataset. Nevertheless, the external validation with the TCGA dataset still preserved the overall trend where more complete distribution of the patch scores provided better predictive ability than the extreme top/bottom instances.

### 3.2 Multiple Neighborhood Instances vs. Single Instance at Each Percentile

Figure 3 clearly shows that multiple instances outperformed the single instance at each percentile. For the cross validation using the MCO CRC dataset, the average C-index was improved from 0.640 to 0.645 when the number of instances at each percentile increased from 1 to 3 (Figure 3a). The average C-index increased to 0.647 with 5 neighborhood instances, and the improvement appears to level off with more neighborhood instances (i.e. 7). Similar pattern was observed with the independent TCGA dataset. The average C-index plateaued (0.580) when the number of instances at each percentile increased to 3. Therefore, consistent with Zhu et al's finding, multiple neighborhood instances at each percentile can improve the model performance. However, experiments may be needed to determine the optimal number of instances for different tasks and models.

### 3.3 Comparison with Baseline Algorithms

Our proposed DeepDistMISL demonstrated superior predictive ability compared to all the baseline algorithms (Figure 4). Compared to MesoNet[8], DeepDistMISL provided an additional 6.3% and 2.8% improvement of mean C-index in the 5-fold cross-validation and external validation, respectively. In addition, DeepDisMISL also outperformed the most recently published, state-of-the-art algorithm, DeepAttnMISL for the MCO CRC dataset [6]. The mean C-index of DeepDisMISL was markedly higher than that of DeepAttnMISL (0.647 vs. 0.606) for the MCO CRC dataset. The superiority of the proposed DeepDistMISL in both cross validation and external validation indicates the robustness of this algorithm and highlights the importance of using complete distribution of patch scores in predicting models. Furthermore, Maxpooling (both top 1 and top 1o instances) had the worst performance compared to other approach in both cross validation and external validation. Similar to the finding in Courtiol et al , although meanfeaturepool provided better performance than MesoNet in cross validation, meanfeaturepool seems less robust and had lower C-index in the external validation (Figure 4b) compared to MesoNet [8]. It is interesting to notice that the simple meanpooling approach had the second-best performance. Meanpooling outperformed all the other baseline approaches in the internal cross validation using the MCO CRC dataset and maintained its performance in the external validation (i.e., provided similar C-index to MesoNet and outperformed other baseline algorithms). It should be mentioned that the results of DeepAttnMISL were obtained directly from the previous publication, and no external validation was conducted for DeepAttnMISL [6].

### 3.4 Risk stratification

Figure 5 shows that DeepDisMISL provided the best risk stratification for both MCO and TCGA populations. Among all the studied deep learning algorithms, the DeepDisMISL provide the most statistically significant separation of the survival curves between the high- and low-risk groups in both MCO and TCGA populations ($p < 0.0001$ in MCO and $p = 0.01$ in TCGA). That is, the DeepDisMISL identified high-risk subgroup presented significantly worse overall survival compared to the low-risk subgroup. MesoNet identified risk groups also showed apparent separation for survival, but with slightly larger p values ($p = 0.001$ in MCO and $p = 0.02$ in TCGA) compared to those based on DeepDisMISL. However, the other baseline algorithms (Meanpooling, Maxpooling with top 1 instance, maxpooling with top 10 instances, and MeanFeaturePool) only provided clear risk stratification in the MCO CRC (training) dataset. In the TCGA COAD READ (external validation) dataset, no statistically significant separation was observed for these baseline algorithms.

### 3.5 Interpreting DeepDisMISL

Figure 7 shows the relationship between the risk and the percentiles of the tile scores. Every percentile appears correlated with risk, i.e., a positive relationship was observed between risk and percentile of tile scores for 0, $0.1^{th}$, $1^{st}$, $5^{th}$, $10^{th}$, and $25^{th}$ percentiles whereas a negative relationship was observed for $50^{th}$, $75^{th}$, $90^{th}$, $99^{th}$, $99.9^{th}$, and $100^{th}$ percentiles. The strong correlation between risk and individual tile score percentiles may explain the decent performance of some published algorithms such as maxpooling and meanpooling. In addition, the tile scores at individual percentiles may carry different information regarding the survival risk. Therefore, combining certain individual percentiles (such as MesoNet were top/bottom 10 instances were used as prediction features) has been shown to provide better predictive performance. This may also explain the excellent performance for DeepDisMISL and illustrate the importance of utilizing the information of the entire distribution (i.e., combining multiple percentiles).

Figure 8 shows the representative tiles near the percentiles. It is interesting to notice that at lower percentiles (e.g., 0 – 75%), the tiles primarily included tumor cells. On the other hand, at the higher percentiles (e.g., 90%, 95, 99%, 99.9%, 100%), muscle appears to be the predominant tissue type. This may partly explain why at lower percentiles, there appears to be positive relationship between risk and the tile scores, whereas a reversed, negative relationship between risk and tile scores was observed at higher percentiles (Figure 7). It is intuitive that the tumor patches are related to higher risk while normal muscle patches may represent lower risk. This suggests that our algorithm is well interpretable and can help to reveal the relationship between morphological phenotypes and patients' risk.

### 3.6 Attention aggregation

We attempted to use attention mechanism in addition to the percentile structure on multiple neighborhood instances i.e., we assigned the same weight to each percentile location [6]. However, the results in Table 1 shows that attention mechanism did not improve the predictive ability. The C-index on MCO with cross validation from the attention-based model was 0.627, while the C-index using external validation dataset TCGA was 0.566. It is worth mentioning that due to large number of parameters required for attention mechanism-based models, overfitting may be a challenge and prevented further improving the prediction.

4. **Discussion**

Risk stratification for cancer patients is currently largely based on cancer staging, mutation/molecular subtyping, and clinical features. Recently, outcome prediction algorithms based on histopathology images using deep learning have been proposed to stratify patients [6, 7, 17-23]. Although difference approaches have been proposed to develop the deep learning survival models, identification of the most predictive image patches is an attractive strategy and has been an active area of research. Patches with highest score are often considered carrying the most predictive value. In addition, Courtiol et al. trained a deep learning model (MesoNet) and demonstrated that adding the patches with the lowest scores to the patches with highest scores can provide excellent prediction of survival in patients with mesothelioma.

In this study, we proposed DeepDisMISL, a patch-score distribution-based multiple-instance survival learning algorithm and demonstrated that other patches also carry important information required to predict survival. We believe this is due to the fact that patches don't convey the complete clinical scenario of a tumor. As such, similar to clinician's algorithms need to leverage all the information available to form a clinical impression. By incrementally adding additional patch scores at different percentiles (e.g., 1st, 5th, 25th, 75th, 95th, and 99th percentiles) to the highest and lowest scoring patches, the predictive performance of DeepDisMISL for survival prediction in colorectal cancer was improved drastically.

We also systematically compared our proposed DeepDisMISL with six existing models: MesoNet [8], DeepAttnMISL [6], Meanpooling, Maxpooling (Top 1 instance), Maxpooling (Top 10 instance) and MeanFeaturePool. DeepDisMISL was not only superior to MesoNet [8], but also outperformed the most recently published state-of-the-art DeepDisMISL. The mean C-index of our proposed DeepDisMISL (0.647) was markedly higher than that of DeepAttnMISL (C-index = 0.606) for the MCO CRC dataset [6]. Compared to MesoNet [8], DeepDisMISL provided 6.3% and 2.8% improvement of mean C-index in the 5-fold cross-validation and external validation, respectively.

The interpretability of deep-learning models is critical and can help to understand the underlying pathology and inform future directions for model improvements. The MesoNet identified image patches were highly interpretable i.e., the high-risk patches were mainly located in stroma regions [8] for patients with mesothelioma. DeepDisMISL was also highly interpretable and revealed a positive relationship between tumor tissues and the risk of death, and a negative relationship between normal muscle tissues and the risk of death. This suggests that DeepDisMISL can help to detect the predictive morphological phenotypes.

It is known that, without external validation, deep learning models are prone to a high risk of bias due to batch effects [24]. As such, we validated and compared our model to other state-of-the-art algorithms using not only 5-fold internal cross validation, but also externally on an independent TCGA dataset. Both validations showed that the proposed DeepDisMISL provided superior performance over all the baseline algorithms, indicative of the robustness of our findings. Further evaluation and applications of DeepDisMISL in other types of cancers and in different population of colorectal cancer patients is warranted.

## 5. Conclusion

We developed DeepDisMISL, a novel distribution based multiple-instance survival learning algorithm to validate our hypothesis of incorporating wholistic patch information within a WSI can predict the CRC cancer survival more accurately. Instead of using just the patches with highest and lowest score, we also used patches that were scored based on the percentile distributions together with the patches that were scored as highest and lowest. We demonstrated that this approach can drastically improve prediction of CRC patient survival outcomes. Including multiple neighborhood instances around each selected distribution location (e.g., percentiles) can further improve the predictive performance of DeepDisMISL. When compared against the six state-of-the art baseline algorithms, DeepDisMISL demonstrated better prediction performance, and more accurate risk stratification for overall survival on both MCO CRC and TCGA COAD-READ datasets. DeepDisMISL is highly interpretable with ability to reveal the relationships and interdependencies between morphological phenotypes and patient's cancer prognosis risk.

**Acknowledgements:**

The research of Xingyu Li, Min Cen, and Hong Zhang are partially supported by National Natural Science Foundation of China (No. 12171451), Anhui Center for Applied Mathematics. Jitendra Jonnagaddala is funded by the Australian National Health and Medical Research Council (No. GNT1192469). Jitendra also acknowledges the funding support received through the Research Technology Services at UNSW Sydney, Google Cloud Research (award# GCP19980904) and NVIDIA Academic Hardware grant programs. We also would like to thank the SREDH Consortium's ([www.sredhconsortium.org](www.sredhconsortium.org)) Translational Cancer Bioinformatics working group for the assistance with the MCO CRC dataset access.

**Contributions:**

X.S.X., X.L., and H.Z. contributed to design of the research; J.J., X.L., and X.S.X. contributed to data acquisition; X.L., X.S.X., and M.C. contributed to data analysis. X.L., X.S.X., J.J., and H.Z. contributed to data interpretation. X.L., X.S.X., J.J., and H.Z. wrote the manuscript; and all authors critically reviewed the manuscript and approved the final version.

**Figure Legends**

**Figure 1.** Pipelines for MesoNet and DeepDisMISL. In multiple instance learning, each data sample is a bag of instances, and the bag can be seen as one patient in our approach. We extract features of tiles from the raw whole slide images, calculat tile-level scores, then obtain percentile scores to predict patient-level risk.

**Figure 2.** Performance for models with different percentiles (single tile at each percentile) for MCO CRC dataset (internal validation) and TCGA dataset (external testing). Black solid dots = mean C-index value from 5-fold cross-validation experiments; Red solid dots = individual C-index values; Error bar = standard deviation.

**Figure 3**. Effects of number of neighborhood tiles at each percentile on model performance for MCO CRC dataset (internal validation) and TCGA dataset (external testing). Solid line = smoothing curve; shaded area = standard deviation around the smoothing line; Black solid dots = mean C-index value from 5-fold cross-validation experiments; Red solid dots = individual C-index values; Error bar = standard deviation.

**Figure 4** Comparison of different baseline algorithms using MCO CRC dataset (internal validation) and TCGA dataset (external testing).

**Figure 5.** Kaplan-Meier plots comparing different algorithms using MCO CRC dataset (internal validation) and TCGA dataset (external testing). For each algorithm, the median risk score in the training set was calculated and then applied as a threshold to stratify each patient into high-risk or low-risk group.

**Figure 6**: Distributions of tile scores. Patients are stratified into 10 different risk groups.

**Figure 7**: Relationship between the average tile score at each percentile and the risk. Patients are stratified into 10 different risk groups.

**Figure 8:** Visualization the spatial locations and morphologic features of tiles near different percentiles in a whole slide image. 13 different colors represent 13 percentile locations. 3

neighborhood tiles at each percentile are selected for visualization, and the tissue type of each tile are predicted by a published CRC tissue classifier.

# Figure 1 (Pipeline)

### (a) DeepDisMISL with single instance at each percentile

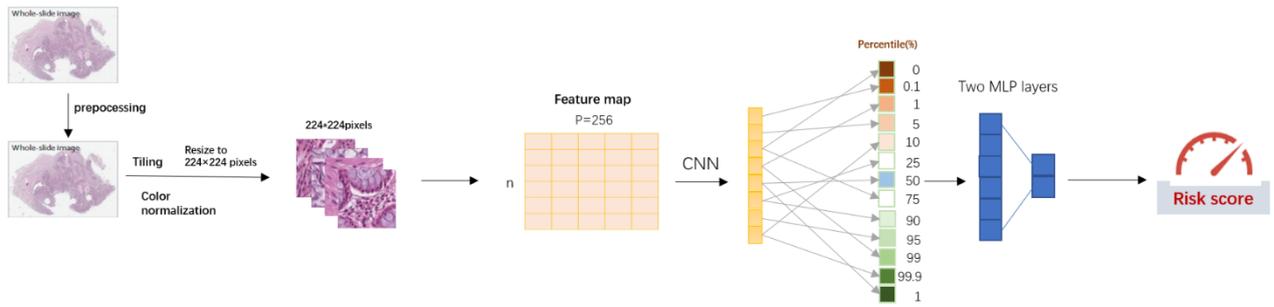

### (b) DeepDisMISL with multiple instances at each percentile:

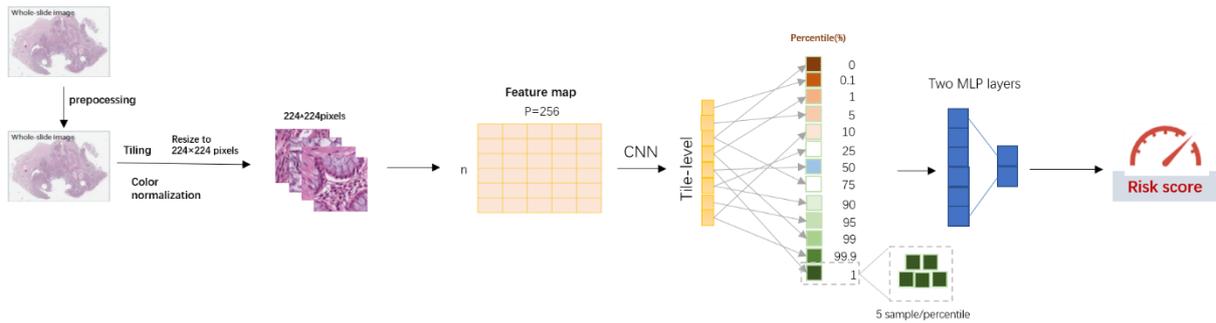

### (c)
### DeepDisMISL with only top and bottom instances:

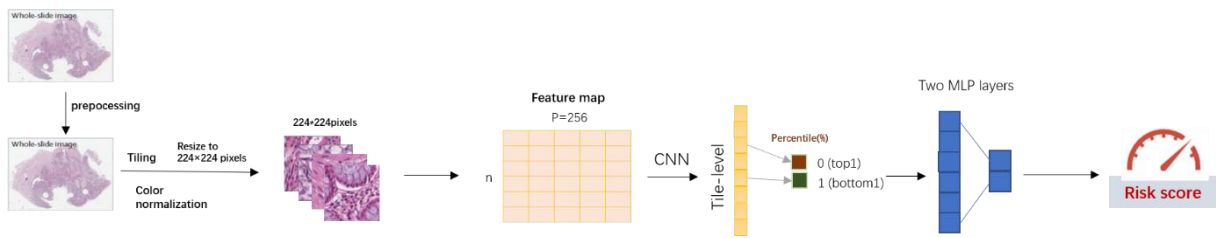

**Figure 2:**

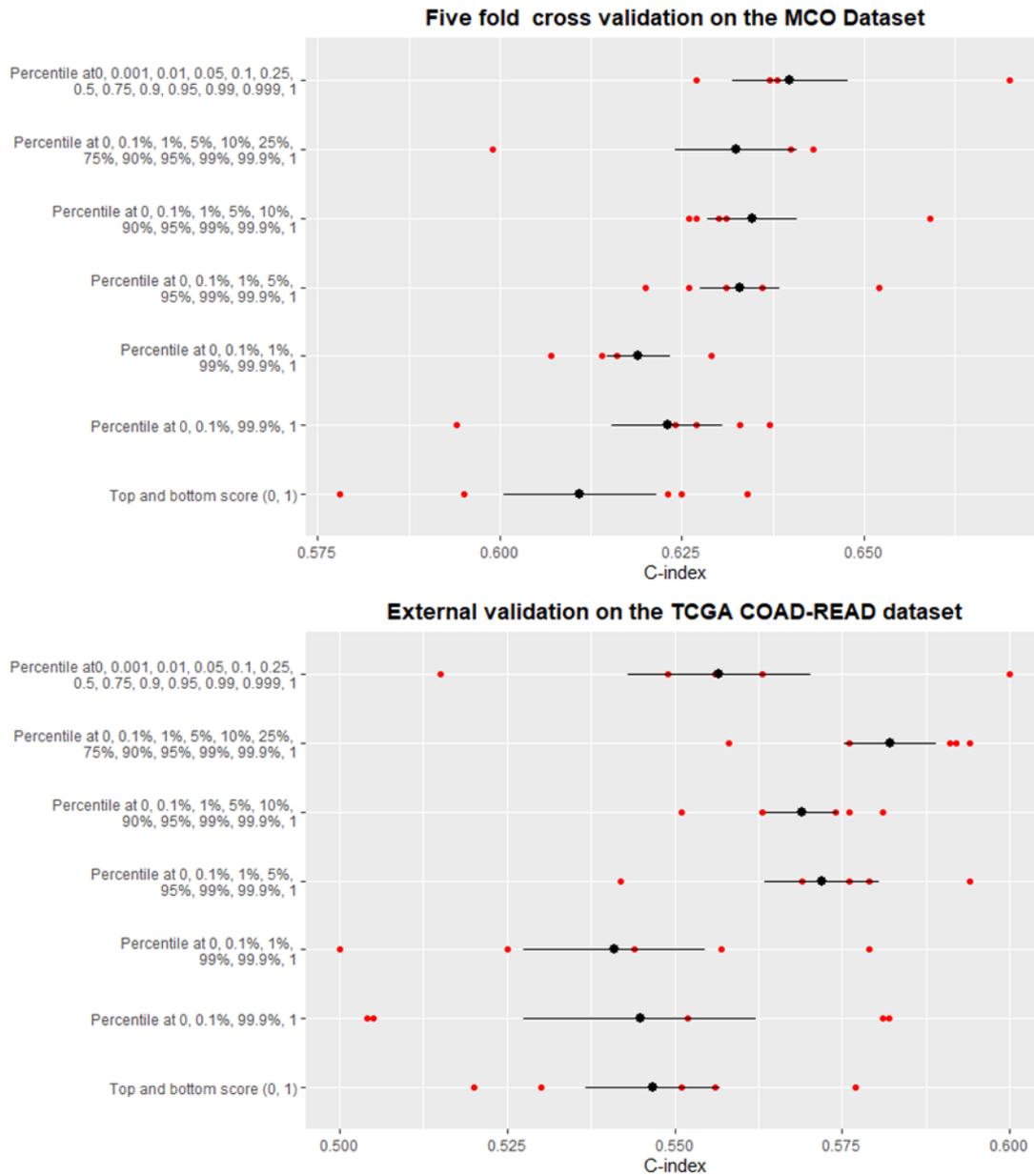

**Figure 3:**

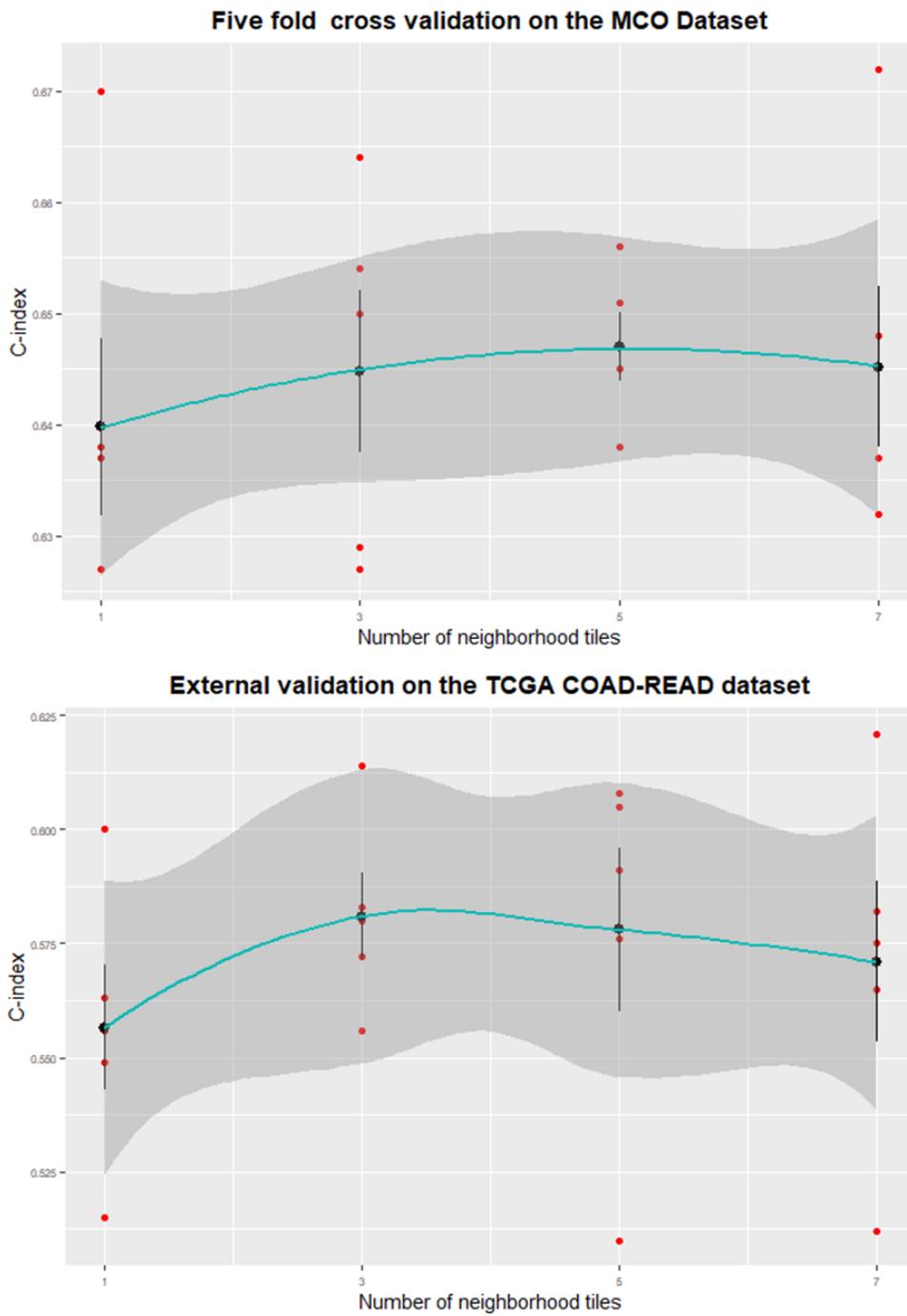

**Figure 4:**

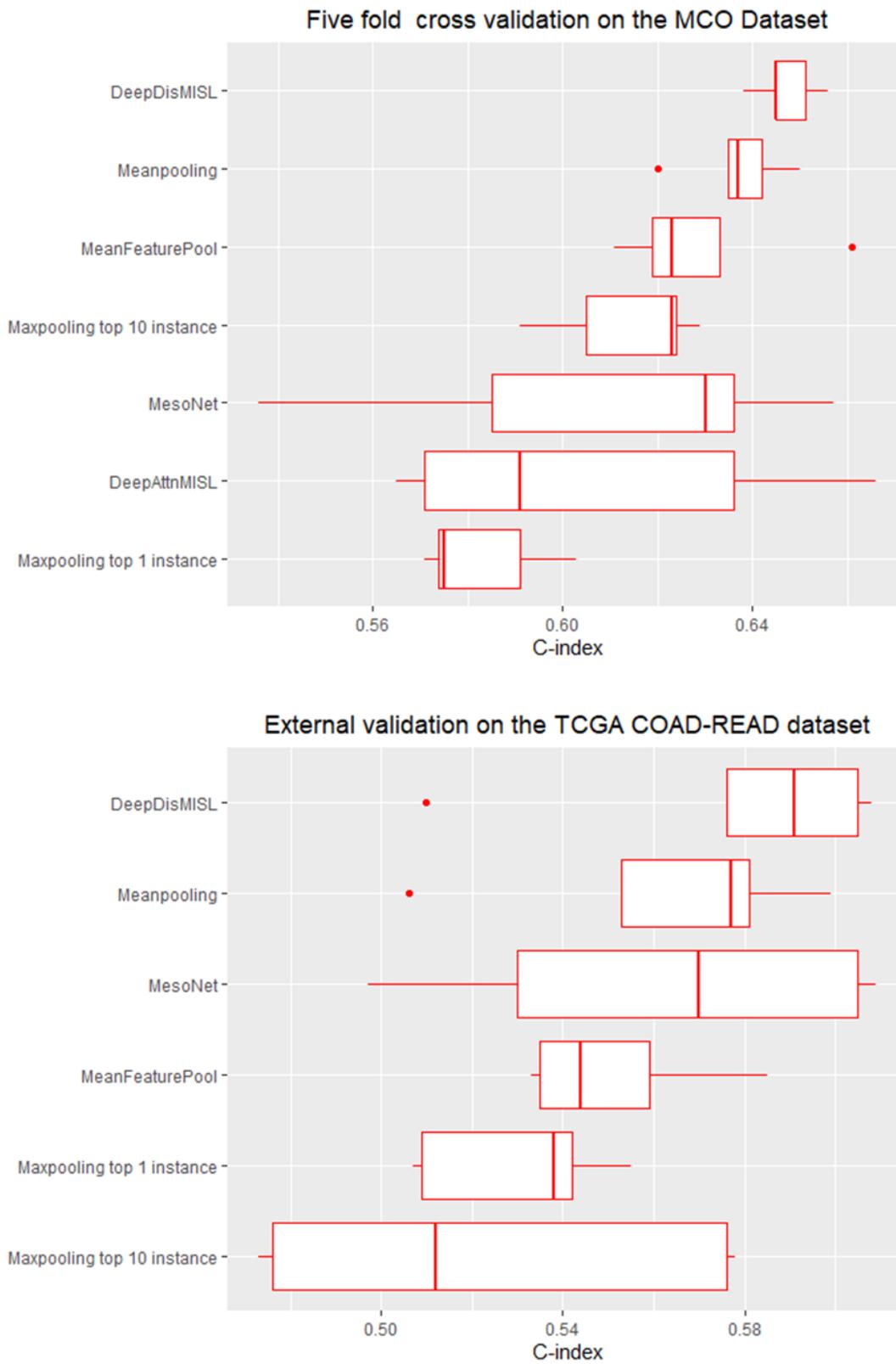

**Figure 5:**

DeepDisMISL                         MesoNet                         Meanpooling

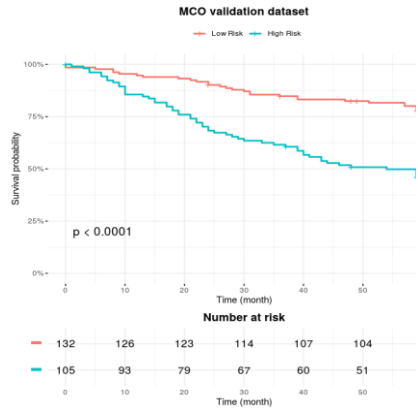 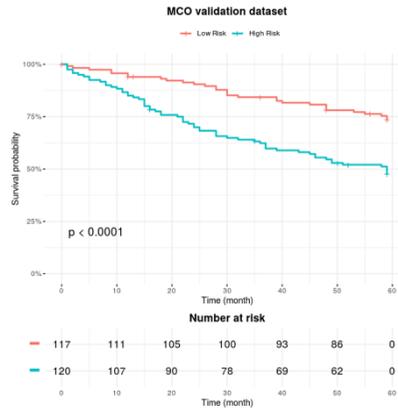 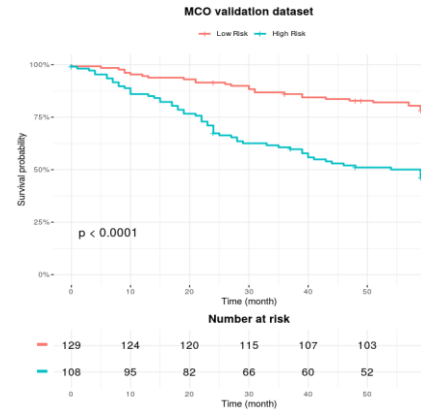

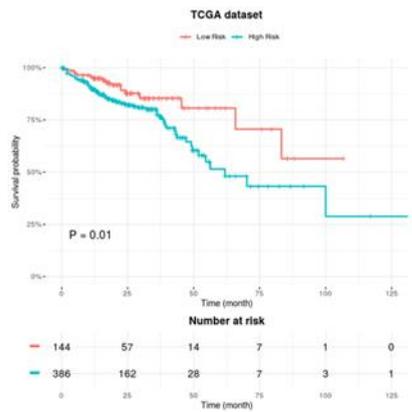 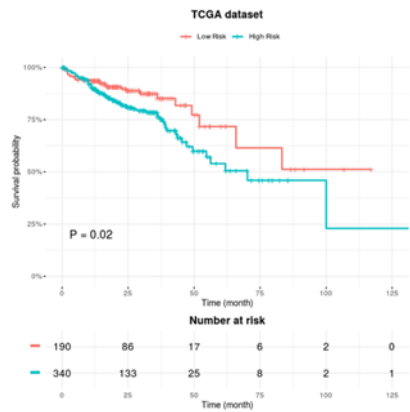 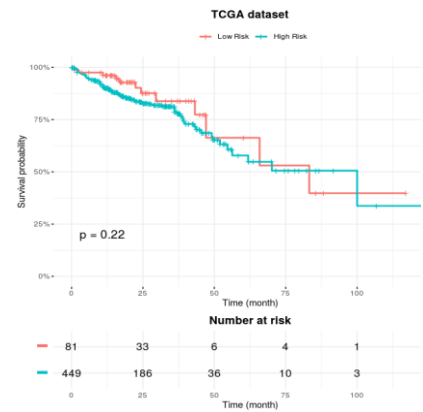

| Top 1 instance | Top 10 instance | MeanFeaturePool |
|---|---|---|
| 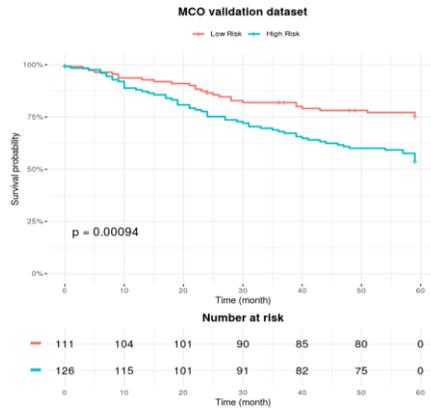 | 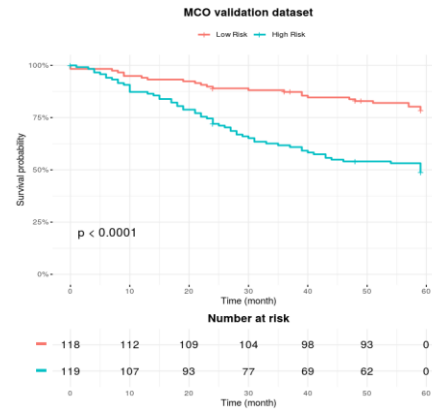 | 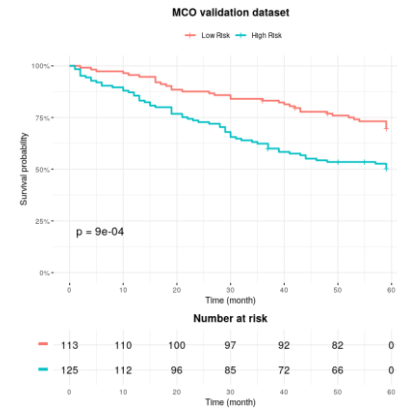 |
| 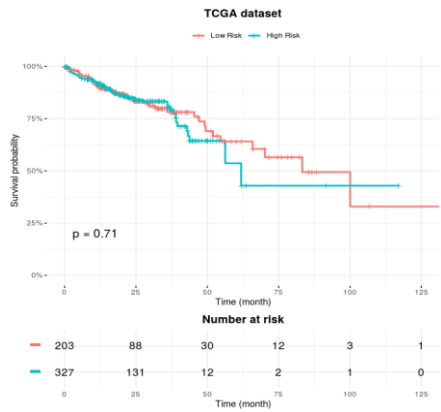 | 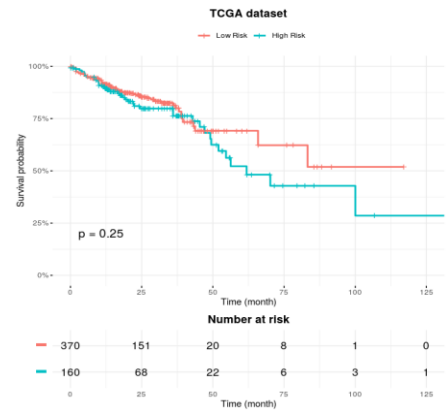 | 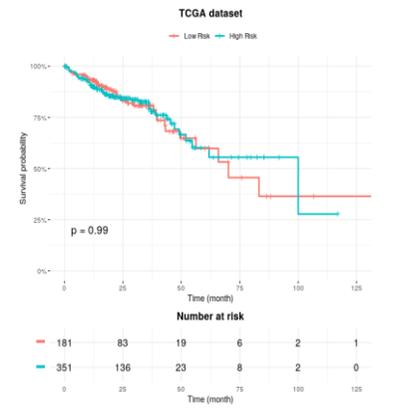 |

Figure 6:

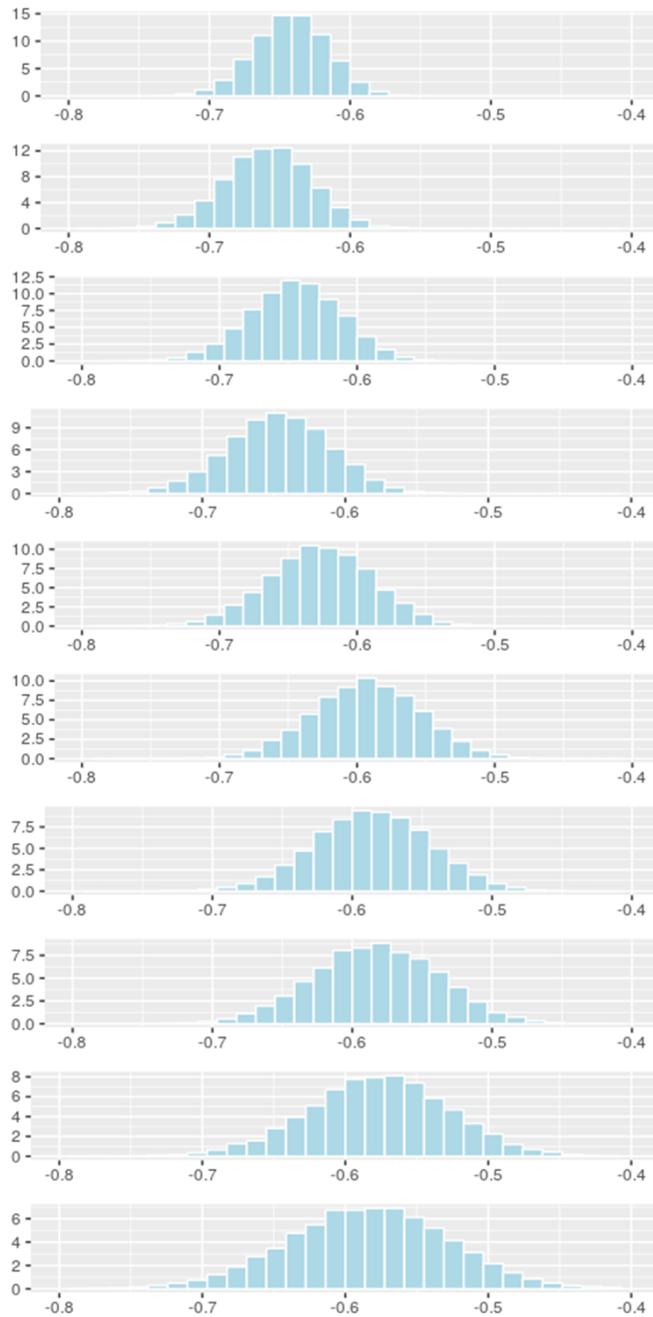

**Figure 7:**

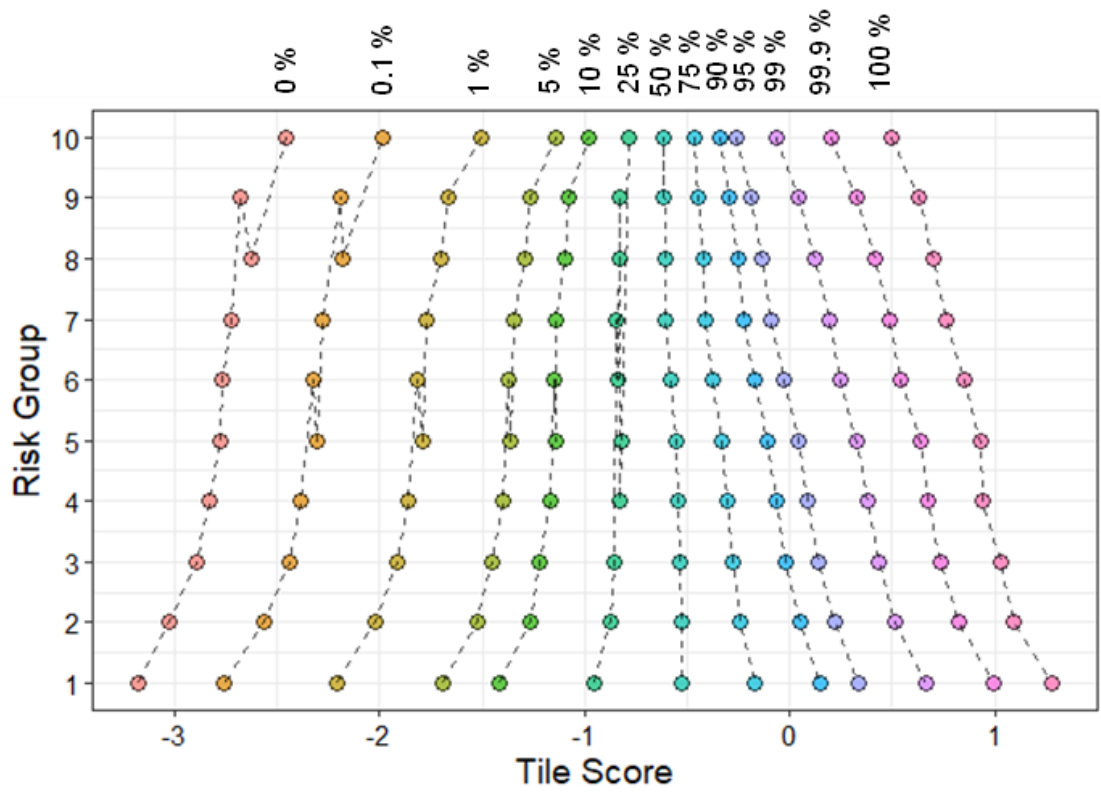

**Figure 8:**

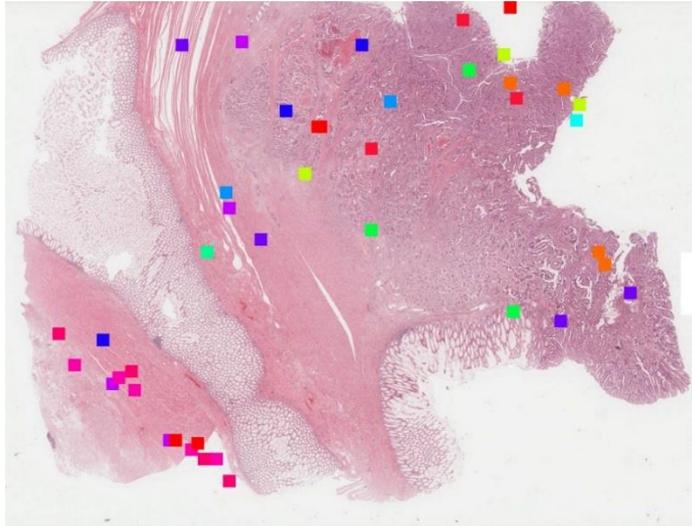

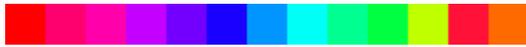

| 0 | STR | TUM | TUM | 0.1% | STR | TUM | TUM |
|---|---|---|---|---|---|---|---|
| 1% | TUM | TUM | TUM | 5% | STR | ADI | TUM |
| 10% | NORM | MUS | STR | 25% | TUM | TUM | TUM |
| 50% | STR | TUM | TUM | 75% | TUM | TUM | TUM |
| 90% | MUS | MUS | MUS | 95% | MUS | MUS | MUS |
| 99% | MUS | MUS | MUS | 99.9% | MUS | MUS | MUS |
| 100% | STR | STR | TUM | | | | |

**Table 1.** Performance of attention mechanism and no attention mechanism.

| MCO CRC dataset - Internal Validation | | | | | |
|---|---|---|---|---|---|
| | Fold 1 | Fold 2 | Fold 3 | Fold 4 | Fold 5 |
| No attention mechanism | 0.66 | 0.64 | 0.65 | 0.65 | 0.65 |
| With attention mechanism | 0.67 | 0.65 | 0.62 | 0.61 | 0.6 |
| TCGA COAD-READ dataset - External Validation | | | | | |
| No attention mechanism | 0.61 | 0.58 | 0.61 | 0.51 | 0.59 |
| With attention mechanism | 0.55 | 0.58 | 0.6 | 0.56 | 0.54 |

Table 2: The structure of the DeepDisMISL

$m_i$: the number of percentile points  $O_i$: risk score

| Layer | Input | Output size |
|---|---|---|
| 1D convolution layer | 12000 × 256 | 12000 × 128 |
| 1D convolution layer | 12000 × 128 | 12000 × 1(score) |
| Aggregating layer | 12000 × 1 | $m_i$ × 1 |
| Fully-Con. | $m_i$ × 1 | 128 × 1 |
| Fully-Con. | 128 × 1 | 1 ($O_i$) |